\documentclass[twocolumn,showpacs,preprintnumbers]{revtex4}
\usepackage{amsmath}
\usepackage{graphicx}
\usepackage{dcolumn}
\usepackage{bm}

\begin{document}

\title{Quantum criticality of the sub-Ohmic spin-boson model within displaced Fock
states}
\author{Shu He$^{1}$, Liwei Duan$^{1}$, and Qing-Hu Chen$^{1,2,*}$}

\address{
$^{1}$ Department of Physics, Zhejiang University, Hangzhou 310027,
P. R. China \\
$^{2}$   Center for Statistical and Theoretical Condensed Matter
Physics, Zhejiang Normal University, Jinhua 321004, P. R. China
 }

 \date{\today }

\begin{abstract}
The spin-boson model is analytically studied using displaced Fock states
(DFS) without discretization of the continuum bath. In the orthogonal
displaced Fock basis, the ground-state wavefunction can be systematically
improved in a controllable way. Interestingly, the zeroth-order DFS
reproduces exactly the well known Silbey-Harris results. In the framework of
the second-order DFS, the magnetization and the entanglement entropy are
exactly calculated. It is found that the magnetic critical exponent $\beta$
is converged to $0.5$ in the whole sub-Ohmic bath regime $0<s<1$, compared
with that by the exactly solvable generalized Silbey-Harris ansatz. It is
strongly suggested that the system with sub-Ohmic bath is always above its
upper critical dimension, in sharp contrast with the previous findings. This
is the first evidence of the violation of the quantum-classical Mapping for $%
1/2<s<1$.
\end{abstract}


\maketitle

The spin-boson model ~\cite{Leggett,weiss} describes a qubit (two-level
system) coupled with a dissipative environment represented by a continuous
bath of bosonic modes. There are currently considerable interests in this
quantum many-body system due to the rich physics of quantum criticality and
decoherence~\cite{weiss,Hur,Kopp}, applied to the emerging field of quantum
computations~\cite{Thorwart}, quantum devices~\cite{mak}, and quantum
biology~\cite{reng,omer}. It is widely used to study the microscopic
behavior of the open quantum systems~\cite{Leggett}. The coupling between
the qubit and the environment is characterized by a spectral function $%
J(\omega )$ which is proportional to $\omega ^s$. The spectral exponent $s$
varies the coupling into three different cases: sub-Ohmic ($s<1$), Ohmic ($%
s=1$), and super-Ohmic ($s>1$).

As a paradigmatic model to study the influence of environment on the quantum
system, the spin-boson model has been extensively and persistently studied
by many analytical and numerical approaches. On the analytical side, a
pioneer work is undoubtedly the variational study based on the polaronic
unitary transformation by Silbey-Harris (SH) ansatz~\cite{Silbey}. Based on
the GHZ ansatz, Zheng \textit{et al.} developed an analytical approach \cite
{zheng} to study both static and dynamical behavior of the dissipative
two-level system. Chin \textit{et al.} generalized the Silbey-Harris (GSH)
variational polaronic ansatz to a asymmetrically one in the sub-Ohmic
spin-boson model~\cite{Chin}. All these studies are based on single coherent
state in both levels. Recently, this single coherent states ansatz was
improved by simply adding other coherent states on the equal footing\cite
{mD1} and by superpositions of two degenerate single coherent states\cite
{ZhengLu}. By the way, the similar idea was also proposed by one of the
present author and a collaborator in 2005 for single-mode case\cite{ren}
independently.

On the numerical side, almost all advanced numerical approaches in the
quantum many-particle physics have been applied and extended to this model.
The numerical renormalization group (NRG) was applied at the earlier stage%
\cite{Bulla} for the sub-Ohmic baths, but the direct applications yields
incorrect critical exponents of the quantum phase transitions (QPT) for $%
0<s<1/2$ and therefore invalidate the famous quantum-to-classical
correspondence due to the Hilbert-space truncation error and the mass flow
error\cite{vojta,phi,Tong}. Later on, quantum Monte Carlo simulations based
on a imaginary path internal~\cite{QMC}, sparse polynomial space approach~%
\cite{ED}, exact diagonalization in terms of shift bosons \cite{Zhang} have
sequentially developed and all found the mean-field critical exponent for $%
0<s<1/2$. The density matrix renormalization group (DMRG)~ was also applied,
but not successful in the analysis of the critical phenomena~\cite{DMRG}.
More recently, using the DMRG algorithm combined with the optimized phonon
basis, a variational matrix product state (MPS) approach formulated on a
Wilson chain\cite{VMPS} was developed and the Hilbert-space truncation can
be alleviated systematically. Very recently, an alternative to the
conventional MPS representation was also proposed\cite{Frenzel}. For $1/2<s<1
$, the magnetic critical exponent $\beta$ obtained in two MPS approaches~%
\cite{VMPS,Frenzel} and the  NRG \cite{Bulla,Tong} is much less than $0.5$,
indicating that the system is below its upper critical dimension.

Among the numerical approaches to the celebrated continuum spin-boson model,
the discretization of the energy spectrum of the bath should be performed at
the very beginning, except for some approaches formulated on path integral%
\cite{QMC} where the bath is analytically integrated out. Whether the
artificial discretization will change the nature of the model system is
still unclear. To ensure the convergence, the number of the bosonic modes
then is set large enough so that the Hilbert-space truncation can be
controlled systematically, and therefore the bath are described in a very
complicated way, like in the various MPS approaches ~\cite{VMPS,Frenzel} and
NRG \cite{Bulla,Tong}. To the best of our knowledge, the phonon state in the
bath of the spin-boson model has not been analytically well described,
except for approaches with more than one nonorthogonal coherent states \cite
{mD1,ZhengLu}.

In this work, we propose an analytic ground state (GS) for the spin-boson
model without discretization of the spectra. The phonon state is expanded in
the novel orthogonal basis, and therefore described in a controllable way.
The GS wavefuction can be obtained self-consistently, and all GS properties
can then be numerically exactly calculated. The convergency of the
criticality is discussed without ambiguity.

The Hamiltonian of the spin-boson model is given by
\begin{equation}
H=-\frac \Delta 2\sigma _x+\sum_k\omega _ka_k^{\dagger }a_k+\frac 12\sigma
_z\sum_kg_k(a_k^{\dagger }+a_k),  \label{hamiltonian}
\end{equation}
where $\sigma _x$ and $\sigma _z$ are Pauli matrices, $\Delta $ is the
tunneling amplitude between two levels, $\omega _k$ and $a_k^{\dagger }$ are
the frequency and creation operator of the $k$-th harmonic oscillator, and $%
g_k$ is the interaction strength between the $k$-th bosonic mode and the
local spin. The spin-boson coupling is characterized by the spectral
function,
\begin{equation}
J(\omega )=\pi \sum_kg_k^2\delta (\omega _k-\omega )=2\pi \lambda \omega
_c^{1-s}\omega ^s,0<\omega <\omega _c,
\end{equation}
with $\omega _c$ a cutoff frequency. The dimensionless parameter $\lambda $
denotes the coupling strength. The rich physics of the quantum dissipation
is second-order QPT from delocalization to localization for $0<s<1$, as a
consequence of the competition between the amplitude of tunneling of the
spin and the effect of the dissipative bath.

To outline the approach more intuitively, we first consider the case without
symmetry breaking, such as the delocalized phase. By using $|\uparrow
\rangle $ and $|\downarrow \rangle $ to represent the eigenstate of $\sigma
_z,$ the GS wavefucntion can be in principle expressed in the following set
of complete orthogonal basis $\prod_{i=0}^na_{k_i}^{\dagger }|0\rangle $

\begin{eqnarray}
&&|\Psi ^{\prime }\rangle =\left( 1+\sum_k\alpha _ka_k^{\dagger
}+\sum_{k_1,k_2}u_{k_1,k_2}a_{k_1}^{\dagger }a_{k_2}^{\dagger }+...\right)
|0\rangle |\uparrow \rangle   \nonumber \\
&&+\left( 1-\sum_k\alpha _ka_k^{\dagger
}+\sum_{k_1,k_2}u_{k_1,k_2}a_{k_1}^{\dagger }a_{k_2}^{\dagger }+...\right)
|0\rangle |\downarrow \rangle ,  \label{Fock}
\end{eqnarray}
where $|0\rangle $ is vacuum of bath modes, $\alpha _k,u_{k_1k_2},...$ are
the coefficients and even parity is considered, However, it is practically
impossible to perform direct diagnializaion in this way to get reasonable
results, because very high order expansions is needed. Alternatively, the
wavefucntion (\ref{Fock}) can be also expressed in terms of the other set of
complete orthogonal basis,$\;D\left( \alpha _k\right) {\prod }%
a_{k_i}^{\dagger }|0\rangle \;$with $D\left( \alpha _k\right) =\exp \left[
\sum_k\alpha _k\left( a_k^{\dagger }-a_k\right) \right] $ a unitary
operators with displacement $\alpha _k$ given in Eq. (\ref{Fock}), as
\begin{eqnarray}
&&|\Psi \rangle =D\left( \alpha _k\right) \left(
1+\sum_{k_1,k_2}b_{k_1k_2}a_{k_1}^{\dagger }a_{k_2}^{\dagger }+...\right)
|0\rangle |\uparrow \rangle   \nonumber \\
&&+D\left( -\alpha _k\right) \left(
1+\sum_{k_1,k_2}b_{k_1k_2}a_{k_1}^{\dagger }a_{k_2}^{\dagger }+...\right)
|0\rangle |\downarrow \rangle ,  \label{DFS}
\end{eqnarray}
where the linear term $a_k^{\dagger }|0\rangle $ should be absent because
the expansion of the whole phonon state of each level in the Fock space can
completely reproduce the first two terms in Eq. (\ref{Fock}). Note above
that the phonon state in each level is generated by operating on the Fock
state with a unitary displacement operators, thus we call it as displaced
Fock states (DFS). Only the first term $D\left( \pm \alpha _k\right)
|0\rangle $ can reach the whole Hilbert-space, so no truncation is made in
this sense. If the expansion is taken to infinity, a exact solution would be
obtained. In other words, the true wavefunction should take the form of Eq. {%
\ref{DFS}) . However, it is impossible to really perform an infinite
expansion. Even for a few terms expansion, it is very time consuming.
Fortunately, it will be shown later that only two terms in the expansion
would give the converging results in some important issues. }

First, as a zeroth-order DFS, we only consider the first term in Eq. (\ref
{DFS}). Projecting the Schr\"{o}dinger equation onto the orthogonal states $%
\left\langle 0\right| \ D^{\dagger }\left( \alpha _k\right) \;$and$\
\left\langle 0\right| a_kD^{\dagger }\left( \alpha _k\right) $ gives
\begin{equation}
\sum_k\omega _k\alpha _k^2+\sum_kg_k\alpha _k-\frac \Delta 2\exp \left[
-2\sum_k\alpha _k^2\right] =E,  \label{Eq_ohmic_01}
\end{equation}
\begin{equation}
\omega _k\alpha _k+\frac 12g_k+\Delta \exp \left[ -2\sum_k\alpha _k^2\right]
\alpha _k=0,  \label{Eq_ohmic_02}
\end{equation}
where we have used the properties of the unitary displacement operators
\begin{equation}
D^{\dagger }\left( \alpha _k\right) a_k^{\dagger }D\left( \alpha _k\right)
=a_k^{\dagger }+\alpha (k);\;D^{\dagger }\left( \alpha _k\right) a_kD\left(
\alpha _k\right) =a_k+\alpha (k).  \nonumber
\end{equation}
Eq. (\ref{Eq_ohmic_02}) immediately yields
\begin{equation}
\alpha _k=\frac{-\frac 12g_k}{\omega _k+\Delta \exp \left( -2\sum_k\alpha
_k^2\right) },  \label{Eq_dis}
\end{equation}
Interestingly, this is just the SH result, although here it is not obtained
through a variational scheme. So we arrive at the right track of the
previous well-known analytical results only by the zeroth-order
approximation. The advantage of the this technique is that we can easily go
further to get more accurate results in a controllable way, by both
modifying the displacement of the unitary operators and adding the
correlations among different bosonic modes step by step.

In the second-order DFS, we only keep two terms in Eq. (\ref{DFS}).
Similarly, projecting the Schr\"{o}dinger equation onto $\left\langle
0\right| \ D^{\dagger }\left( \alpha _k\right) ,\;\left\langle 0\right|
a_kD^{\dagger }\left( \alpha _k\right) $ , and $\ \left\langle 0\right|
a_{k_1}a_{k_2}D^{\dagger }\left( \alpha _k\right) \;$yields the following
three equations for unknown $E,\alpha _k$, and$\ b_{k_1,k_2}$,
\begin{equation}
E=\sum_k\left( \omega _k\alpha _k^2+g_k\alpha _k\right) -\frac 12\Delta \eta
\left( 1+4\sum_kB_k\alpha _k\right) ,\   \label{Eq_2_en}
\end{equation}
\begin{equation}
\alpha _k=-\frac{\frac{g_k}2+2\sum_{k^{\prime }}b_{k,k^{\prime }}\left[
\left( \omega _{k^{\prime }}-\Delta \eta \right) \alpha _{k^{\prime }}+\frac{%
g_{k^{\prime }}}2\right] }{\omega _k+\Delta \eta \left( 1+4\sum_kB_k\alpha
_k\right) },  \label{Eq_21}
\end{equation}
\begin{equation}
b_{k_1,k_2}=-\frac{B_{k_1}\alpha _{k_2}+B_{k_2}\alpha _{k_1}-\alpha
_{k_1}\alpha _{k_2}\left( 1+4\sum_kB_k\alpha _k\right) }{2\sum_kB_k\alpha
_k+\left( \omega _{k_1}+\omega _{k_2}\right) /\left( \Delta \eta \right) },
\label{Eq_22}
\end{equation}
where
\begin{eqnarray*}
B_k &=&\sum_{k^{\prime }}b_{k,k^{\prime }}\alpha _{k^{\prime }}, \\
\eta  &=&\exp \left[ -2\sum_k\alpha _k^2\right] .
\end{eqnarray*}
Both $\alpha (k)$ and$\ b_{k_1,k_2}$ can be obtained by solving the two
coupled equations (\ref{Eq_21}) and (\ref{Eq_22}) self-consistently, which
in turn give the GS energy and wavefunction. In our opinion, this is
actually a parameter-free analytical approach.

Due to the QPT from the delocalized phase to the localized one in the
sub-Ohmic spin-boson model, we should relax wavefucntions (\ref{DFS}) to the
asymmetrical one
\begin{eqnarray}
&&|\Psi \rangle =D\left( \alpha _k\right) \left(
1+\sum_{k_1,k_2}b_1(k_1,k_2)a_{k_1}^{\dagger }a_{k_2}^{\dagger }+...\right)
|0\rangle |\uparrow   \nonumber \\
&&+D(\beta _k)\left( r+\sum_{k_1,k_2}b_2(k_1,k_2)a_{k_1}^{\dagger
}a_{k_2}^{\dagger }+...\right) |0\rangle |\downarrow \rangle ,
\label{sub-Ohmic}
\end{eqnarray}
where $r$ is the asymmetrical parameter. If $r=1$ and $\beta _k=-$ $\alpha _k
$, the previous symmetrical results are recovered.

The zero-order DFS will give the same results as that in generalized SH
polaronic ansatz \cite{Chin}, then it is also called the GSH ansatz in the
remaining of the paper. In the second-order DFS, we have double equations
for the counterparts in the symmetrical case. The number of unknown
parameters are also doubled, due to the asymmetrical coefficients. We leave
detailed derivations to Appendix A.

Proceeding as the scheme outlined above, we can straightforwardly perform
the further expansion in the orthogonal diplaced Fock basis $D\left( \alpha
_k\right) {\prod }a_{k_i}^{\dagger }|0\rangle $ in the controllable way, and
obtain the solution within any desired accuracy in principle. The challenges
remain on the pathway to high dimensional integral in the further
extensions, due to both the analytical derivations and exponentially
increasing computational difficulties. On the other hand, the criterion of
the precise description of the criticality can be that the further
correction does not change the nature in the last approximation.
Fortunately, it will be shown later that the second-order correction really
does not changes the critical exponents in the GSH ansatz at all, so the
further corrections to the second-odder DFS is not necessary, at least in
the sense of the criticality.

The magnetization $\langle \sigma _z\rangle $ can be used as an order
parameter in the QPT of this model. It shows a power law behavior near the
critical point,
\begin{equation}
\langle \sigma _z\rangle \propto \left( \lambda -\lambda _c\right) ^\beta .
\end{equation}
The entanglement entropy between the qubit and the bath is defined as \cite
{Osterloh}
\begin{eqnarray*}
S &=&-Tr\rho _A\log _2\rho _A=-Tr\rho _B\log _2\rho _B, \\
\rho _{A(B)} &=&Tr_{B(A)}\left\langle \Psi \right| \Psi \rangle ,
\end{eqnarray*}
where $A$ is the qubit and $B$ is the bath, $\Psi $ is the GS wavefucntion
of the whole system. In the spin-boson model , it is \cite{Kopp}
\[
S=-p_{+}\log _2p_{+}-p_{-}\log _2p_{-},
\]
where
\[
p_{\pm }=\left( 1\pm \sqrt{\langle \sigma _x\rangle ^2+\langle \sigma
_z\rangle ^2}\right) /2.
\]

We stress here that in the present approach we do not need to discreatize
the bosonic energy band like in many previous studies at the very beginning.
All $k-$summation in the coupled equations can be transformed into
continuous $\ $integral like$\;\int_0^{\omega _c}d\omega J(\omega )I(\omega
) $. In this work, all integrals are numerically calculated within a
Gaussian-logarithmical (GL) integration  with very high accuracy. The detailed
demonstration is given in Appendix B. Without loss of generality, we set $%
\Delta =0.1,\omega _c=1$ in the calculation throughout this paper, if not
specified.

First, we calculate the magnetization and evaluate the critical points
within both the GSH and the second-order DFS. The results for $s=0.2,0.4,0.6$%
, and $0.8$ are presented in Fig. 1 (a) by the solid lines (second-order
DFS) and dashed lines (GSH). Both shows that there exist a critical point
which separate the delocalized phase ($\langle \sigma _z\rangle =0$) to the
localized one ($\langle \sigma _z\rangle \neq 0$). The critical coupling
strengths$\;\lambda _c$ by the second-oeder DFS are larger than those by
GSH, the correction becomes remarkable for $s>0.5$. It follows that the GSH
critical point will be modified by the second-order DFS. To determine the
critical point within the second-order DFS more precisely, we also calculate
the entanglement entropy, which exhibits a cusp characteristics around the
critical point. The results are given in Fig. 1(b). Both the magnetization
and the entanglers entropy result in the consistent value for the QPT
critical point.

\begin{figure}[tbp]
\includegraphics[width=8cm]{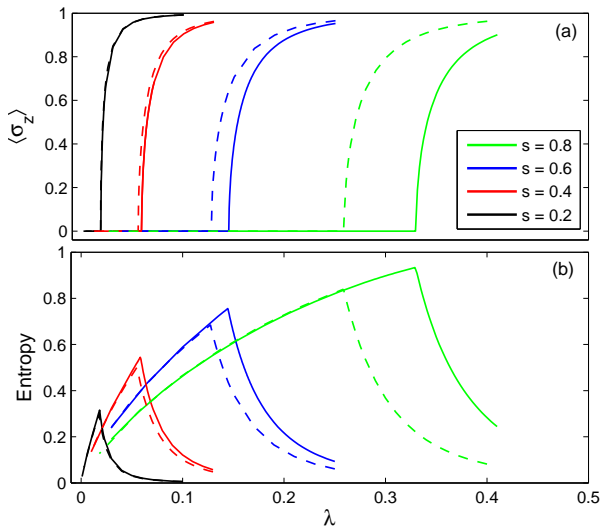}
\caption{(Color online) The magnetization and the entanglement entropy as a
function the coupling strength within the GSH (dashed lines) and the second order DFS (solid lines) for $%
s=0.2,0.4,0.6$, and $0.8$.}
\label{tunneling}
\end{figure}

It should be pointed out that that the critical coupling strength obtained
in the second-order DFS is not the true one in this model either, because it
will be definitely revised by the third-order DFS, although the revision is
probably small. It is expected that the converging critical point would be
only obtained in the high-order DFS, which is however a challenging task at
the moment, and also beyond the scope of the present study. The more crucial
issue in a QPT is the criticality. So the natural question is '' the
criticality described by GSH could be changed in second-order DFS?''

In Fig. 2, we present the magnetization within both GSH and the second-order
DFS as a function of $\lambda /(\lambda -\lambda _c)$ in a log-log plot for $%
s=0.2,0.4,0.6$, and $0.8$. It is demonstrated that, using both approaches,
the magnetic critical exponent$\;\beta $ is always $0.5$ with an error bar $%
(-0.01,0.01)$, even for $s=0.6$ and $0.8$. Note that the second-order DFS
should be the dominate correction to the GSH, as indicted in the critical
points. But for the critical exponent, we do not find any visible deviation
from the GSH ones. We can not imagine that the further corrections would
change this observation but the second-order correction does not. It is
therefore strongly suggested that even in $s>1/2$, the magnetic exponent $%
\beta$ in the sub-Ohmic spin-boson model is always $0.5$, quite different
from those obtained in the MPS \cite{VMPS,Frenzel} and NRG \cite{Bulla,Tong}.

\begin{figure}[tbp]
\includegraphics[width=8cm]{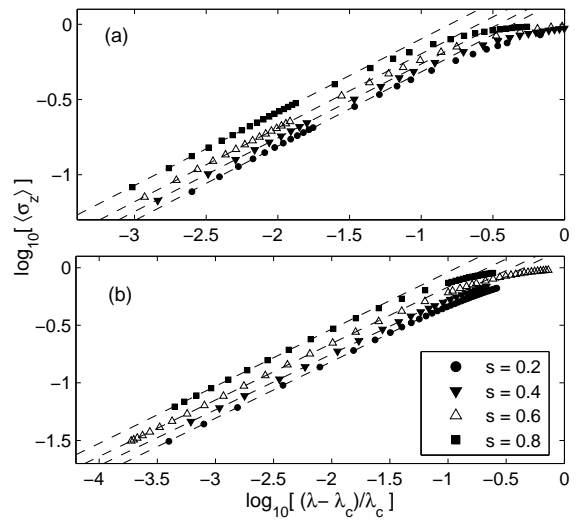}
\caption{ The log-log plot of the magnetization $\langle \sigma _z\rangle $
as a function the coupling strength within the GSH and the second order DFS
for $s=0.2,0.4,0.6$, and $0.8$. $\omega _c=1,\Delta =0.1$ }
\label{critical_curves_ECS}
\end{figure}

In summary, a new analytic approach referred to DFS is proposed in the
spin-boson model with the continuum spectral function. The zero-order
approximation is just the well known SH approach, the further corrections
can be performed step by step. For the sub-Ohmic baths, the second-order DFS
can modify the GSH critical coupling strength of the QPT, especially for $%
s>1/2$. But the critical exponent is not changed at all, and is always $0.5$
for the whole bath regime $0<s<1$, a mean-field value for the system above
the upper critical dimension. This is a direct strict evidence of failure of
quantum-classical mapping in the sub-ohmic spin-boson model, at least for $%
s>1/2$.

\textsl{Outlook}. It is expected that the sufficient number of integral
grids for the converged results increases exponentially with the further
corrections in the DFS. The Monte Carlo integral might be used in the high
dimensional integral. But the analytical derivation in the high order DFS is
also challenging task. New methods, probably like some diagrammatic
techniques, in the framework of the DFS is highly called for. The progress
along this avenue may hopefully lead to a true exact solution to this
celebrated model, which is perhaps our future ambitions.

Because each summation over $k$ in the final expressions is related to $%
\sum_kg_k^2$, we propose a discretized spin-boson Hamiltonian ($\omega _c=1$%
) as follows
\begin{equation}
H=-\frac \Delta 2\sigma _x+\sum_k\omega _ka_k^{\dagger }a_k+\frac 12\sigma
_z\sum_k\sqrt{\frac{W(\omega _k)J(\omega _k)}\pi }(a_k^{\dagger }+a_k),
\label{dis_H}
\end{equation}
where $\omega _k=\omega _{m,n}$ is the Gaussian integration point in Eq. (%
\ref{combination}), $W(\omega _k)$ is the Gaussian weight. Applying the
present DFS approach to this Hamiltonian, all results obtained in this paper
are recovered completely by direct summation over $k$. It is shown in
Appendix B that a limited number of discretizations can give results with
very high accuracy. In this sense, Hamiltonian (\ref{dis_H}) is equivalent
to the model for one qubit coupled with a finite number bosonic modes, which
facilities the further study. We believe that Eq. (\ref{dis_H}) with
discretized bosonic modes could be a new starting Hamiltonian for any
advanced approaches. The dynamics based on polaron trial state by the name
of the Davydov D1 ansatz within the Dirac-Frenkel time dependent variational
procedure \cite{duan} can be revisited using the discretized one directly.

This work is supported by National Natural Science Foundation of China under
Grant No. 11474256, and National Basic Research Program of China under Grant
No.~2011CBA00103.

$^{\ast }$ Corresponding author. Email:qhchen@zju.edu.cn

\appendix

\section{DFS for the sub-Ohmic baths}

In the zeroth order approximation, we only select the first term Eq. (\ref
{sub-Ohmic}). Similar to the derivation in the symmetric case, projecting
the Schr\"{o}dinger equation in the upper level onto the orthogonal basis $%
\left\langle 0\right| \ D^{\dagger }\left( \alpha _k\right) \;$and$\
\left\langle 0\right| a_kD^{\dagger }\left( \alpha _k\right) $ and low level
onto $\left\langle 0\right| \ D^{\dagger }\left( \beta _k\right) \;$and$\
\left\langle 0\right| a_kD^{\dagger }\left( \beta _k\right) $ result in
\begin{eqnarray}
\sum_k\left( \omega _k\alpha _k^2+g_k\alpha _k\right) -\frac \Delta 2r\
\Gamma  &=&E,  \label{E_upper} \\
\omega _k\alpha _k+\frac 12g_k+\frac \Delta 2r\Gamma D_k &=&0,
\label{dis_upper}
\end{eqnarray}
and
\begin{eqnarray}
\sum_k\left( \omega _k\beta _k^2-g_k\beta _k\right) -\frac \Delta {2r}\
\Gamma  &=&E,  \label{E_down} \\
\omega _k\beta _k-\frac 12g_k-\frac \Delta {2r}\Gamma D_k &=&0,
\label{dis_down}
\end{eqnarray}
where
\begin{eqnarray*}
\Gamma  &=&\exp \left[ -\frac 12\sum_kD_k^2\right] , \\
D_k &=&\alpha _k-\beta _k,
\end{eqnarray*}
which are the same as those obtained variationally within the GSH ansatz\
\cite{Chin}.

For the second-order DFS, the first two terms in Eq. (\ref{sub-Ohmic}) is
kept. Proceeding as procedures outlines above, Projecting the
Schr\"{o}dinger equation in the upper level onto the orthogonal states $%
\left\langle 0\right| \ D^{\dagger }\left( \alpha _k\right) ,\ \left\langle
0\right| a_kD^{\dagger }\left( \alpha _k\right) $, and $\ \left\langle
0\right| a_{k_1}a_{k_2}D^{\dagger }\left( \alpha _k\right) $ and low level
onto $\left\langle 0\right| \ D^{\dagger }\left( \beta _k\right)
,\left\langle 0\right| a_kD^{\dagger }\left( \beta _k\right)$, and $\
\left\langle 0\right| a_{k_1}a_{k_2}D^{\dagger }\left( \beta _k\right) $
yield the following six equations
\begin{widetext}
\begin{eqnarray}
\sum_k\left[ \omega _k\alpha _k^2+g_k\alpha _k\right] -\frac \Delta 2\Gamma
\left[ r+\sum_kB_kD_k\right] \ &=&E,  \label{sub_1up} \\
r\sum_k\left[ \omega _k\beta _k^2-g_k\beta _k\right] -\frac \Delta 2\Gamma
\left[ 1+\sum_kA_kD_k\right] \ &=&rE,  \label{sub_1down}
\end{eqnarray}

\begin{equation}
\left[ \omega _k\alpha _k+\frac{g_k}2\right] +\sum_{k^{\prime }}2b_1\left(
k,k^{\prime }\right) \left[ \omega _{k^{\prime }}\alpha _{k^{\prime }}+\frac{%
g_{k^{\prime }}}2\right] -\Delta \Gamma B_k+\frac \Delta 2\Gamma D_k\left[
r+\sum_kB_kD_k\right] =0,  \label{sub_2up}
\end{equation}
\begin{equation}
r\left[ \omega _k\beta _k-\frac{g_k}2\right] +\sum_{k^{\prime }}2b_2\left(
k,k^{\prime }\right) \left[ \omega _{k^{\prime }}\beta _{k^{\prime }}-\frac{%
g_{k^{\prime }}}2\right] +\Delta \Gamma A_k-\frac \Delta 2\Gamma D_k\left[
1+\sum_kA_kD_k\right] =0,  \label{sub_2down}
\end{equation}
\begin{eqnarray}
&&b_1(k_1,k_2)\left( \omega _{k_1}+\omega _{k_2}\right) +\frac \Delta
2\Gamma \left[ r+\sum_kB_kD_k\right] b_1(k_1,k_2)  \nonumber \\
&&-\frac \Delta 2b_2(k_1,k_2)\Gamma +\frac \Delta 2\Gamma \left[
B_{k_1}D_{k_2}+B_{k_2}D_{k_1}\right] -\frac \Delta 4\Gamma
D_{k_1}D_{k_2}\left[ r+\sum_kB_kD_k\right] =0,  \label{sub_3up}
\end{eqnarray}
\begin{eqnarray}
&&b_2(k_1,k_2)\left( \omega _{k_1}+\omega _{k_2}\right) +\frac \Delta
{2r}\Gamma \left[ 1+\sum_kA_kD_k\right] b_2(k_1,k_2)  \nonumber \\
&&-\frac \Delta 2b_1(k_1,k_2)\Gamma +\frac \Delta 2\Gamma \left[
A_{k_1}D_{k_2}+A_{k_2}D_{k_1}\right] -\frac \Delta 4\Gamma
D_{k_1}D_{k_2}\left[ 1+\sum_kA_kD_k\right] =0,  \label{sub_3down}
\end{eqnarray}
\end{widetext}
where
\begin{eqnarray*}
\ \ A_k &=&\sum_{k^{\prime }}b_1(k^{\prime },k)D_{k^{\prime }}, \\
\ \ B_k &=&\sum_{k^{\prime }}b_2(k^{\prime },k)D_{k^{\prime }}.
\end{eqnarray*}
The self-consistent solutions for the four coupled equations Eqs. (\ref
{sub_2up}), (\ref{sub_2down}), (\ref{sub_3up}) and (\ref{sub_3down}) will
give all results in the second-order DFS. If set $r=1,\alpha _k=-\beta _k$
and $b_1(k_1,k_2)=b_2(k_1,k_2)$, Eqs. (\ref{Eq_21}) and (\ref{Eq_22}) in the
symmetric case are recovered completely.

\section{Gaussian-logarithmical integration for the continuous integral}

The symmetrical case is also used to illustrate a effective numerical
approach to the calculation of the summation clearly. In the zeroth-order
approximation, also the well known SH ansatz, we can set
\[
\alpha _k=\alpha _k^{\prime }g_k,
\]
Eq. (\ref{Eq_dis}) becomes
\[
\alpha _k^{\prime }=-\frac{1/2}{\omega _k+\Delta \exp \left( -2\sum_k\alpha
_k^{\prime 2}g_k^2\right) },
\]
so $\alpha _k^{\prime }$ is only related to $g_k$ implicitly.

According to the spectral density, we have
\begin{equation}
\alpha ^{\prime }(\omega )=-\frac{1/2}{\omega +\Delta \exp \left[ -\frac
2\pi \int_0^{\omega _c}d\omega ^{\prime }\alpha ^{\prime 2}(\omega ^{\prime
})J(\omega ^{\prime })\right] },  \label{SH_cont}
\end{equation}
which can be solved numerically by iterations.

In the second-order approximation, we can set
\begin{eqnarray*}
\alpha _k &=&\alpha _k^{\prime }g_k, \\
b_{k_1,k_2} &=&b_{k_1,k_2}^{\prime }g_{_{k_1}}g_{_{k_2}}.
\end{eqnarray*}
Inserting to Eqs. (\ref{Eq_21}) and (\ref{Eq_22}) gives
\[
\alpha _k^{\prime }=\frac{-\frac 12+2\sum_{k^{\prime }}g_{k^{\prime
}}^2b_{k,k^{\prime }}^{\prime }\left[ \left( \omega _{k^{\prime }}-\Delta
\eta \right) \alpha _{k^{\prime }}^{\prime }+1/2\right] }{\omega _k+\Delta
\eta \left( 1+4\zeta \right) },
\]

\[
b_{k_1,k_2}^{\prime }=\frac{\alpha _{k_1}^{\prime }\alpha _{k_2}^{\prime
}\left( 1+4\zeta \right) -\sum_{k^{\prime }}g_{k^{\prime }}^2\alpha
_{k^{\prime }}^{\prime }\left( b_{k_1,k^{\prime }}^{\prime }\alpha
_{k_2}^{\prime }+b_{k_2,k^{\prime }}^{\prime }\alpha _{k_1}^{\prime }\right)
}{2\zeta +\left( \omega _{k_1}+\omega _{k_2}\right) /\left( \Delta \eta
\right) },
\]
where
\[
\zeta =\sum_kg_k^2\sum_{k^{\prime }}g_{k^{\prime }}^2b_{k,k^{\prime
}}^{\prime }\alpha _{k^{\prime }}^{\prime }\alpha _k^{\prime }.
\]
Given $g_k$, both $\alpha _k^{\prime }$ and $b_{k_1,k_2}^{\prime }\;$can be
obtained self-consistently. Note that each $k$-summation takes the form of $%
\sum_kg_k^2I(k)$ where $I(k)$ does not depend on $g_k^2\;$explicitly, and so
both $\alpha _k^{\prime }$ and $b_{k_1,k_2}^{\prime }\;$are functionals of $%
g_k$. $\;$Without loss of generality, $k$ is corresponding to $\omega $ one
by one, the $k$-summation can be transformed to the $\omega \;$integral as
\[
\sum_kg_k^2I(k)\rightarrow \int_0^{\omega _c}d\omega \frac{J(\omega )}\pi
I(\omega ),
\]
so we have
\begin{equation}
\alpha ^{\prime }(\omega )=\frac{-\frac 12+\xi (\omega )-2\Delta \eta \chi
(\omega )}{\omega +\Delta \eta \left( 1+4\zeta \right) },
\label{displacement}
\end{equation}

\begin{equation}
b^{\prime }\left( \omega _1,\omega _2\right) =\frac{\alpha ^{\prime }(\omega
_1)\alpha ^{\prime }(\omega _2)\left( 1+4\zeta \right) -\kappa (\omega
_1,\omega _2)}{2\zeta +\left( \omega _{_1}+\omega _{_2}\right) /(\Delta \eta
)},  \label{sec_coeff}
\end{equation}
where
\begin{eqnarray*}
\xi (\omega ) &=&\int_0^{\omega _c}d\omega ^{\prime }\frac{J(\omega ^{\prime
})}\pi \left[ 2\omega ^{\prime }\alpha ^{\prime }(\omega ^{\prime
})+1\right] b^{\prime }\left( \omega ,\omega ^{\prime }\right) , \\
\chi (\omega ) &=&\int_0^{\omega _c}d\omega ^{\prime }\frac{J(\omega
^{\prime })}\pi \alpha ^{\prime }(\omega ^{\prime })b^{\prime }\left( \omega
,\omega ^{\prime }\right) , \\
\kappa (\omega _1,\omega _2) &=&\chi (\omega _1)\alpha ^{\prime }(\omega
_2)+\chi (\omega _2)\alpha ^{\prime }(\omega _1),
\end{eqnarray*}
are some functions for $\omega ,$ and
\begin{eqnarray*}
\zeta  &=&\int_0^{\omega _c}d\omega \frac{J(\omega )}\pi \int_0^{\omega
_c}d\omega ^{\prime }\frac{J(\omega ^{\prime })}\pi \alpha ^{\prime }(\omega
)\alpha ^{\prime }(\omega ^{\prime })b^{\prime }\left( \omega ,\omega
^{\prime }\right) , \\
\eta  &=&\exp \left[ -2\int_0^{\omega _c}d\omega ^{\prime }\frac{J(\omega
^{\prime })}\pi \alpha ^{\prime 2}(\omega ^{\prime })\right] ,
\end{eqnarray*}
are constants. If both $\alpha ^{\prime }(\omega )$ and $b^{\prime }\left(
\omega _1,\omega _2\right) $ are obtained, all observables can in turn be
calculated. For example, using Eq. (\ref{Eq_2_en}), the energy in the
second-order DFS can be calculate as
\begin{equation}
E=\int_0^{\omega _c}d\omega \frac{J(\omega )}\pi \alpha ^{\prime }(\omega
)\left[ \omega \alpha ^{\prime }(\omega )+1\right] -\frac 12\Delta \eta
\left( 1+4\zeta \right) \ .
\end{equation}

The self-consistent solutions in the coupled equations Eqs. (\ref
{displacement}) and (\ref{sec_coeff})) are in no way obtained analytically,
numerical calculation should be performed. Note that the low frequency modes
play the dominant role in the QPT of the sub-ohmic spin-boson model. At the
critical point, there is an infrared divergence of the integrand like $%
\int_0^{\omega _c}\omega ^{s-2}d\omega \;$in the limit of $\omega
\rightarrow 0$ for sub-ohmic bath, which is called as the infrared
catastrophe. Thanks to the Gaussian quadrature rules, where the zero
frequency is not touched. We can discretize the whole frequency interval
with Gaussian grids, the integral can be numerically exactly achieved with a
large number of Gaussian grids. It is very time consuming to calculate the
integral in this way, especially for high dimensional integral involved in
the high-order DFS. According to the structure of the integrand, it is not
economical to deal with the high and low frequency regime on the equal
footing. To increase the efficiency, we combine the logarithmic
discretization and Gaussian quadrature rule. First, we divide the $\omega $
interval $[0,1]$ into $M+1$ sub-intervals as $[\Lambda ^{-(m+1)},\Lambda
^{-m}]\;(m=0,1,2,M-1)\;$and$\;[0,\Lambda ^{-M}]$ , then we apply the
Gaussian quadrature rule to each logarithmical sub-interval. So the
continuous integral is calculated by the following summation
\begin{equation}
\int_0^1J(\omega )I(\omega )d\omega =\sum_{m=0}^M\sum_{n=1}^NW_{m,n}J(\omega
_{m,n})I(\omega _{m,n}),  \label{combination}
\end{equation}
where $N$ is the number of gaussian points inserted in each sub-interval, $%
W_{m,n}$ is corresponding Gaussian weight.

\begin{figure}[tbp]
\includegraphics[width=8cm]{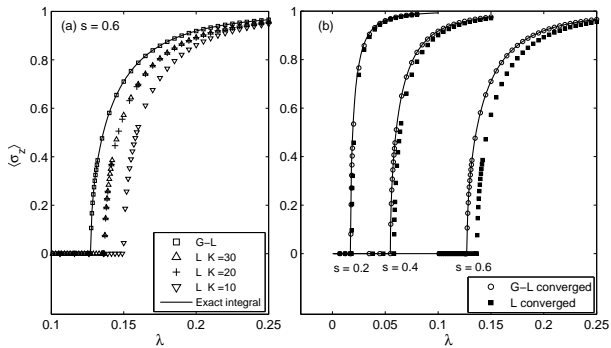}
\caption{ Magnetization $\langle \sigma _z\rangle$ as a function of the
coupling strength $\lambda$ in the GSH ansatz. (a) For $s = 0.6$, converged
results within GL integration (open squares), numerical exact ones (solid
lines), and those within logarithmic discretization with different
truncation numbers $K = 10, 20$, and $30$. (b) The converged magnetization
within GL integration (open circles) and logarithmic discretization(filled squares)  for $s
= 0.2,0.4$, and $0.6$. Numerical exact ones are denoted by the solid lines.}
\label{CompareIntegral}
\end{figure}

\begin{figure}[tbp]
\includegraphics[width=8cm]{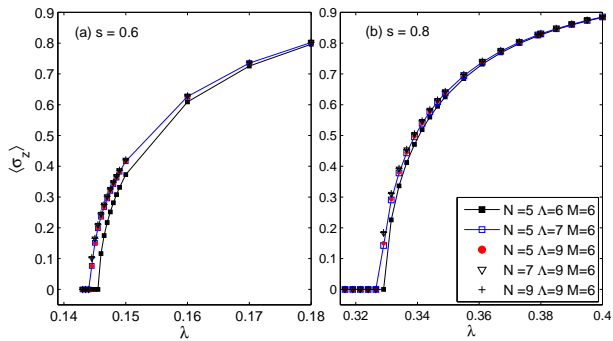}
\caption{ (Color online) Magnetization $\langle \sigma _z\rangle$ as a function of the
coupling strength $\lambda$ in the second-order DFS within GL integration
using different $M, N$, and $\Lambda$ for (a) $s=0.6$ and (b) $s=0.8$(b).}
\label{Fig_converge}
\end{figure}

To demonstrate the efficiency of the Gaussian-logarithmical (GL)
integration, we first apply it to the GSH ansatz, which is also the
zero-order approximation in the DFS. The one-dimensional integral can be
numerically exactly done by Gaussian integration over the whole interval
with a huge number of discretizations, and corresponding results can be
regarded as a benchmark. After careful examinations, using the GL technique,
the converging results for the magnetization can be archived if set $M=6,N=9,
$ and $\Lambda =9$ The corresponding results for $s=0.6$ are presented in
Fig. \ref{CompareIntegral} (a) with open squares, which agrees excellently
with the numerically exact one by a huge number of discretization in the
Gaussian integration (solid lines).

We can also perform the logarithmic discretization of the bosonic energy
band, as was widely used in the previous studies, such as NRG \cite{Bulla}
and multi-coherent states \cite{mD1}. In the GSH ansatz, this can be easily
done by set
\begin{equation}
g_k^2=\int_{\Lambda ^{-(k+1)}}^{\Lambda ^{-k}}\frac{J(\omega )}\pi d\omega
,\;\omega _k=\frac 1{g_k^2}\int_{\Lambda ^{-(k+1)}}^{\Lambda ^{-k}}\frac{%
J(\omega )}\pi \omega d\omega ,  \label{Dis_GSH_H}
\end{equation}
in Eqs. (A1-A4) of Appendix A. The spectral density is truncated to a number
$K$ of modes. The summation is performed over the integer $k$ directly and
the self-consistent solution with discretized form can be also obtained. The
logarithmic grid is chosen as $\Lambda =2$, the same as that in Refs. \cite
{Bulla,mD1}. The magnetization as a function of $\lambda $ for $s=0.6$ with
such a logarithmic discretization are collected in Fig. \ref{CompareIntegral}
(a) with different truncation number $K$ of bosonic modes. The converging
results can be also obtained for $K\geq 20$, which is however obviously
different from the numerically exact one. Note that this kind of logarithmic
discretization of the bosonic energy band at the very beginning is not
equivalent to the logarithmic discretization of the continuous integral
derived in the end of the DFS approach.

The converged magnetization within both the GL and logarithmic
discreatization for different values of $s$ are carefully examined, and the
results are exhibited in Fig. \ref{CompareIntegral} (b). The deviation
between these two convergent ones increases with $s$, and becomes remarkable
for $s\ge 0.3$.

Then we turn to the second-order DFS study, a central issue in this work.
Two-dimensional integral will be involved in this case, so direct Gaussian
integral with huge number of discretization is practically difficult.
Fortunately, it has been convincingly shown above that in the framework of
GSH ansatz, the Gaussian-logarithmic discretization with dozens of grids to
the continuous integral can effectively give results with very high
accuracy. Therefore we extend this numerical technique to the present case.
Interestingly, a excellent convergence behavior for $s=0.6$(a) and $s=0.8$%
(b) is demonstrated in Fig. \ref{Fig_converge} with different value of $M,N$%
, and $\Lambda $. The converged results obtained in this way compose the main achievement in this
work.

\end{document}